\documentclass[acmsmall,screen]{acmart}
\setcopyright{none}
\renewcommand\footnotetextcopyrightpermission[1]{}

\AtBeginDocument{%
  }

\setcopyright{none}
\acmConference[ISSTA 2026]{International Symposium on Software Testing and Analysis}{2026}{}
\acmYear{2026}

\usepackage[most]{tcolorbox}
\usepackage{multirow}
\usepackage{booktabs}
\usepackage{soul}
\usepackage{colortbl}
\usepackage{graphicx}
\usepackage{xcolor}
\usepackage{tcolorbox}
\usepackage{arydshln}
\usepackage{subcaption}
\usepackage{wrapfig}
\tcbuselibrary{listings, breakable}
\tcbset{
  highlightbox/.style={
    colback=gray!10,
    colframe=black!15,
    boxrule=0pt,
    arc=1mm,
    left=1mm,
    right=1mm,
    top=0.5mm,
    bottom=0.5mm,
    enhanced,
    sharp corners,
  }
}

\definecolor{sectiongray}{RGB}{235,235,235}
\definecolor{titleblue}{RGB}{44, 62, 80}

\begin{document}


\title{From Context to Intent: Reasoning‑Guided Function‑Level Code Completion}

\author{Yanzhou Li}
\affiliation{%
  \institution{Nanyang Technological University}
  \country{Singapore}
}
\email{yanzhou.li@ntu.edu.sg}

\author{Tianlin Li}
\affiliation{%
  \institution{Beihang University}
  \country{China}
}
\email{Tianlin001@buaa.edu.cn}

\author{Yiran Zhang}
\affiliation{%
  \institution{Nanyang Technological University}
  \country{Singapore}
}
\email{yiran002@ntu.edu.sg}

\author{Shangqing Liu}
\affiliation{%
  \institution{Nanjing University}
  \country{China}
}
\email{shangqingliu@nju.edu.cn}

\author{Aishan Liu}
\affiliation{%
  \institution{Beihang University}
  \country{China}
}
\email{liuaishan@buaa.edu.cn}

\author{Xianglong Liu}
\affiliation{%
  \institution{Beihang University}
  \country{China}
}
\email{xlliu@buaa.edu.cn}

\author{Yang Liu}
\affiliation{%
  \institution{Nanyang Technological University}
  \country{Singapore}
}
\email{yangliu@ntu.edu.sg}
\renewcommand{\shortauthors}{Trovato et al.}
\newcommand{\lyz}[1]{\textit{\color{orange}#1}}
\newcommand{\ltl}[1]{\textit{\color{blue}#1}}
\newcommand{\yiran}[1]{\textit{\color{cyan}#1}}

\begin{abstract}

The growing capabilities of Large Language Models (LLMs) have led to their widespread adoption for function completion within code repositories. Recent studies on such tasks show promising results when explicit instructions, often in the form of docstrings, are available to guide the completion. However, in real-world scenarios, clear docstrings are frequently absent. Under such conditions, LLMs typically fail to produce accurate completions.
To enable more automated and accurate function completion in such settings, we aim to enable LLMs to accurately infer the developer’s intent prior to code completion. Our key insight is that the preceding code, namely the code context before the function to be completed, often contains valuable cues that help the model understand the intended functionality. However, inferring intent from such implicit context is non-trivial and constitutes a core challenge in function-level code completion. To tackle this challenge, inspired by how humans interpret context, we propose a reasoning-based prompting framework that guides LLMs to utilize these contextual cues to infer intent step by step.
To incentivize LLMs to reason through the preceding code and infer intent, we further curate a dataset of 40k examples, each annotated with intermediate reasoning traces and corresponding docstrings. Extensive experiments on DevEval and ComplexCodeEval demonstrate consistent performance improvements across multiple models, achieving over 25\% relative gains in pass@1 for both DeepSeekCoder and CodeLLaMA families.
Building upon our framework, we further develop an intent-interactive platform that supports lightweight human feedback. This platform allows developers to select from a set of candidate intentions or edit the intent to better guide the model. Our experiments show that this interactive approach leads to further performance improvements.

\end{abstract}

\maketitle

\section{Introduction}
\label{sec:intro}

Recent advances in large language models (LLMs) have led to their widespread adoption in software engineering. 
Tools such as GitHub Copilot and Cursor~\cite{wermelinger2023copilot} integrate LLMs into development environments to assist with tasks like code completion~\cite{jiang2024completionsurvey}, generation~\cite{chen2021humaneval,lu1codexglue}, and bug repair~\cite{jiang2023repair}. 
Among these tasks, code completion is the most widely used and popular feature, with millions of developers relying on it daily to speed up coding \cite{businessofapps_copilot2025}. 


Recent research has moved beyond line-level prediction toward generating larger semantic units such as code blocks and complete function bodies, often in a repository-level setting~\cite{jiang2024completionsurvey,van2023completionmodel,liu2020multicompletion,izadi2024completionevaluation,nguyen2019function,feng2024complexcodeeval}. To address long-range dependencies and cross-file references, many approaches adopt RAG (retrieval-augmented generation), retrieving relevant definitions or usage examples from the same project and incorporating them into the model input~\cite{zhang2023repocoder,ding2023crosscodeeval,cheng2024draco}. While this strategy expands the available context, it implicitly assumes that providing more related code is sufficient to guide accurate generation. In practice, however, function-level completion frequently remains unreliable even with rich contextual information, because the model lacks a clear and explicit representation of the developer’s intent. 
As a result, the primary challenge is not only the availability of contextual information, but also the absence of effective intention guidance during the completion process, which has received limited attention in prior studies on function-level and repository-level code completion.


While recent studies on code completion tasks show promising results when clear docstrings are provided to explicitly convey the developer’s intent~\cite{zhang2023repocoder,feng2024complexcodeeval,li2024deveval},  
models exhibit a substantial performance gap when the developer’s intent is unclear or missing, compared to settings where accurate intent is explicitly specified. To further quantify this gap, we conduct a preliminary study (detailed in Section~\ref{sec:gap}).  
In practice, with the increasing automation of software development workflows, developers often expect models to complete function-level code even without explicit guidance. Meanwhile, writing comprehensive docstrings is time-consuming and frequently neglected, and lightweight or convention-based documentation in real-world projects often fails to convey precise intent. Therefore, addressing code completion in the absence of clear intentions represents an important and underexplored problem in practical development settings.

This performance gap largely stems from the fact that, in the absence of explicit intent, the function signature alone often provides limited insight into the function’s purpose. Our key insight is that preceding code within the same file, such as key variables, control structures, and helper functions, often contains valuable clues about the intended functionality. Therefore, we aim to leverage this contextual information to better infer the developer’s intent. However, we find that LLMs still struggle to accurately extract functional intent from preceding code for precise completion. Even when explicitly prompted to infer intent from context, LLMs often fail to produce satisfactory results, revealing inherent limitations in their ability to reason about and synthesize intent from surrounding code.\looseness=-1

To address this challenge, we draw inspiration from how humans interpret context and propose a prompting framework that guides LLMs to reason through the preceding code to infer intent. Specifically, our prompt template consists of three components. The first component instructs the model to perform lexical inference by extracting cues from elements such as the file name, function name, and argument names. The second component guides the model to conduct a semantic analysis of the preceding code, summarizing the existing functionality and identifying what remains to be implemented. Finally, the third component prompts the model to synthesize these insights into a descriptive function docstring that explicitly captures the inferred intent.
To equip LLMs with structured reasoning capabilities for intention inference, we curate annotated reasoning trajectories and adopt supervised fine-tuning to internalize this process. This enables the model to autonomously perform intent reasoning from context during inference.
Specifically, we construct a fine-tuning dataset consisting of paired inputs (prompts generated by our reasoning framework) and corresponding outputs (reasoning traces and docstrings). We begin by manually annotating a small set of seed examples and then use GPT-4o~\cite{hurst2024gpt} to expand the dataset via few-shot in-context generation. This results in approximately 40,000 training instances. We use this data to fine-tune models from the CodeLLaMA~\cite{roziere2023codellama} and DeepSeekCoder~\cite{zhu2024deepseekcoder} families, equipping them with the ability to infer developer intent through structured reasoning prior to code generation.

Moreover, building upon our reasoning-based framework, we introduce an optional three-stage pipeline for intent-aware code completion. In practical settings, this pipeline preserves a fully automated and efficient default workflow, while allowing developers to optionally enter a lightweight interaction loop for improved controllability and closer alignment with their actual intentions. Specifically, the model first generates a small set of candidate intentions through lightweight reasoning. Developers can then optionally select or refine one of the candidates to better reflect their actual goal. Finally, the model completes the function based on the selected or edited intention.



To evaluate the effectiveness of our approach for code completion, we conduct experiments on two repository-level benchmarks: DevEval~\cite{li2024deveval,li2024evocodebench} and ComplexCodeEval~\cite{feng2024complexcodeeval}. We assess performance using both reference-based metrics, including CodeBLEU and Edit Similarity~\cite{ren2020codebleu}, and execution-based metrics such as pass@1~\cite{chen2021humaneval}. Our results show that intention inference through explicit reasoning consistently improves performance over baseline models. Across different model sizes, we observe absolute gains exceeding 5\% in both pass@1 and reference-based metrics, with relative improvements surpassing 25\% on pass@1, demonstrating the substantial effectiveness of our approach.
Under simulated lightweight human interactions for quantitative evaluation, intent-interaction yields additional improvements of over 2\% in both Pass@1 and reference-based metrics. Our framework significantly narrows the performance gap to the oracle setting with access to ground-truth intentions, reducing the Pass@1 difference from over 15\% to within 4\% comparing to the baselines.


In summary, this paper makes the following key contributions:

\begin{itemize}
  \item We reveal that current LLMs often fail to accurately infer the intention of the target function, which leads to poor performance.
  
  \item We propose a reasoning-based framework to infer intent step by step and incentivize code LLMs to reason about intent through supervised fine-tuning.

  \item We introduce a three-stage intent-interactive framework that incorporates optional and lightweight developer interaction to further improve completion effectiveness.
  
  \item We conduct comprehensive experiments demonstrating the importance of intention inference and structured reasoning, and show that our method significantly improves function-level code completion.
\end{itemize}

\section{Background}
\label{sec:background}

\subsection{Problem Definition}


Developers increasingly rely on AI-powered coding assistants, such as Copilot and Cursor, to streamline software development. One highly anticipated capability is function-level code completion, which aims to generate an entire function body based on surrounding code context.
In practice, users rarely provide complete or accurate docstrings, yet still expect models to infer the intended functionality and produce coherent implementations.
In this work, we study the task of function-level code completion, where the objective is to generate a complete function body given only its surrounding in-file context, including the file name, preceding code, and function signature.

Specifically,
Given a file \( C \) with name \( C_{\text{name}} \), we define a target function \( f \in C \) consisting of a signature \( f_{\text{sig}} \) and a body \( f_{\text{body}} \). The code preceding \( f \) is denoted as \( C_{\text{pre}} \subset C \).
The goal of function-level code completion is to generate the function body given all available context using a language model \( M \):
$$
\hat{f}_{\text{body}} = M\left( C_{\text{name}}, C_{\text{pre}}, f_{\text{sig}} \right).
$$

\subsection{Motivation}
\begin{figure}[t]
    \centering
    \includegraphics[width=0.9\linewidth]{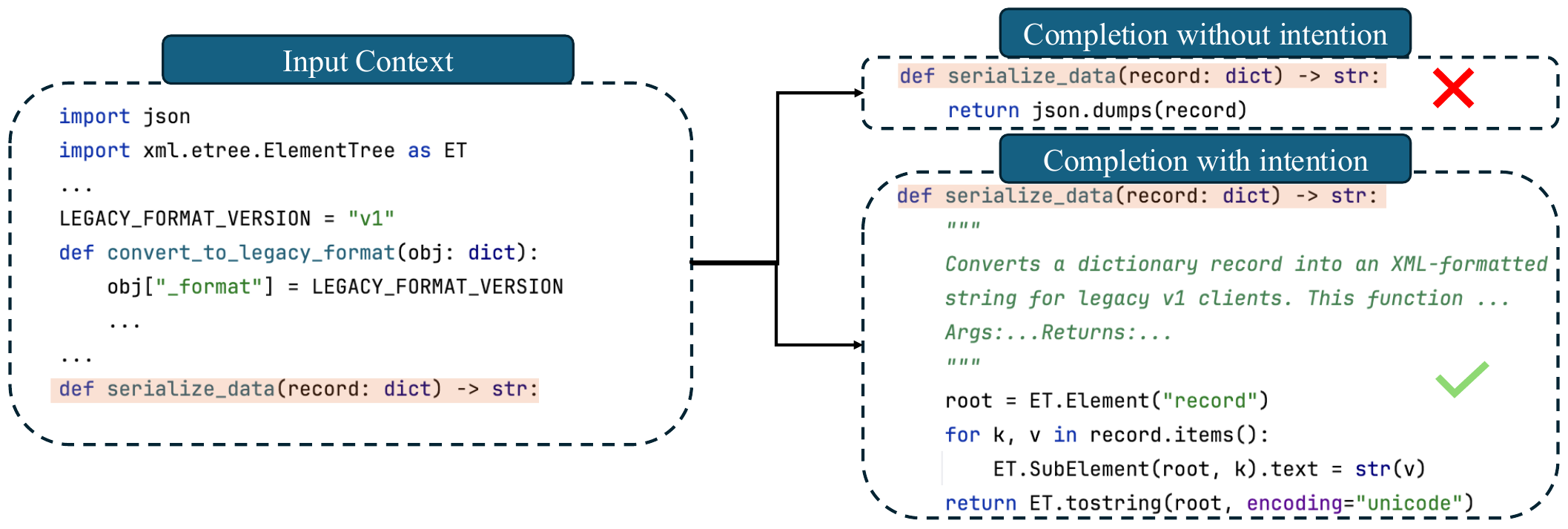}
    \caption{
    A motivation example
    }
    \label{fig:motivation}
    \vspace{-8pt}
\end{figure}

\label{sec:gap}

When there are no clear instructions, function-level code completion within code repositories presents a unique challenge: the model must infer what the target function is intended to accomplish based solely on the preceding context.
While cues such as the function name, signature, and preceding code may implicitly suggest the intended functionality, it remains unclear whether the model can effectively leverage these indirect cues to infer the high-level intent of the function it is completing.

To better understand this, we compare model performance with and without access to the ground-truth docstring, which serves as an explicit representation of the function’s intended purpose. 
If a model can already infer the correct intention from context, the addition of the docstring should lead to only marginal improvements.
We here analyze a sample as shown in Figure \ref{fig:motivation}, when completing the function serialize\_data, the model without access to the docstring produces a generic implementation that serializes the input to JSON. The model here completes the function with surface-level pattern recognition that overlooks a subtle but important clue in the surrounding context. Specifically, the presence of the v1 version tag implies that the output should be formatted as XML in order to maintain compatibility with legacy systems. In contrast, when the ground-truth docstring is provided, the model correctly generates XML serialization logic. This example illustrates that even strong code LLMs may fail to accurately infer function-level intent from preceding context, especially when key signals are implicit. 
input

The experimental results are summarized in Table~\ref{tab:motivation}. Across all three models, we observe a consistent and substantial gap in function-level code completion performance between the setting without a docstring and the setting where the ground-truth docstring is provided. Specifically, all models show relative improvements exceeding 40\% in pass@1 accuracy when given the correct docstring. This indicates that having access to an explicit description of the function’s intention significantly improves generation quality. The consistency of this pattern suggests that current models often fail to reliably infer the intended functionality from the surrounding context alone. As a result, completing functions without such explicit guidance remains highly error-prone. 

These findings motivate us to develop a framework in which the model first infers the target function’s intention in a structured and automated manner—aligned with the developer’s actual development goals—before generating the implementation.
To achieve this, we examine how intent is implicitly embedded in real-world codebases. In particular, lexical cues such as file names, function signatures, and argument names often hint at the function’s role. Additionally, the preceding code, especially helper functions and control logic, provides rich semantic signals about the broader functionality.
Inspired by how developers structure and understand code, we design a multi-step reasoning framework that guides the model to extract and synthesize these contextual cues into a coherent intent representation before code generation.

\begin{table}[t]
\caption{Pass@1 accuracy of function-level code completion on the DevEval dataset}
\label{tab:motivation}
\centering
\small
\begin{tabular}{lccc}
\toprule
\textbf{Model} & \textbf{w/o Docstring} & \textbf{w/ Docstring} & \textbf{Relative $\uparrow$} \\
\midrule
CodeLLaMA-7B        & 17.51\% & 32.29\% & +84.4\% \\
DeepSeekCoder (16B)  & 21.29\% & 34.27\% & +60.97\% \\
GPT-4o              & 29.56\% & 42.62\% & +44.18\% \\
\bottomrule
\end{tabular}
\vspace{-6pt}
\end{table}

\section{Methodology}
\label{sec:method}
To address the intention inference gap identified in Section~\ref{sec:gap}, we propose a two-part framework designed to improve function-level code completion. First, we enhance the model’s ability to infer the intention of a target function by explicitly generating it prior to code generation. To improve the accuracy and context sensitivity of this process, we introduce a structured step-by-step reasoning procedure inspired by chain-of-thought reasoning~\cite{wei2022cot}. Second, we extend this capability with a three-stage intent-interactive workflow that incorporates lightweight user input. It allows users to optionally select or refine candidate intentions before generation, improving alignment with user goals while maintaining the efficiency of the overall completion pipeline.

\subsection{Reasoning-Based Prompting Framework}

As shown in the previous section, LLMs often struggle to infer the intention of a target function from its surrounding context.
From a developer’s perspective, however, code is typically structured to reflect high-level goals. Even in the absence of explicit docstrings, cues such as the function signature and surrounding helper functions often imply what the target function is meant to accomplish.
This suggests that code context carries rich yet implicit signals of intent.
Nonetheless, extracting these signals requires interpreting lexical elements, analyzing what the preceding code achieves, and identifying what functionality remains to be implemented.

To address this, we explicitly formulate intention inference as a step-by-step reasoning task. Inspired by recent work on model reasoning and chain-of-thought prompting~\cite{wei2022cot,li2025scodecot,diao2024cot2}, we design a fixed reasoning template to guide the model in analyzing contextual cues and incrementally inferring functionality.

\textbf{Intention Reasoning Prompt Template.}
\label{sec:cot}
\begin{figure}[t]
    \centering
    \includegraphics[width=0.9\linewidth]{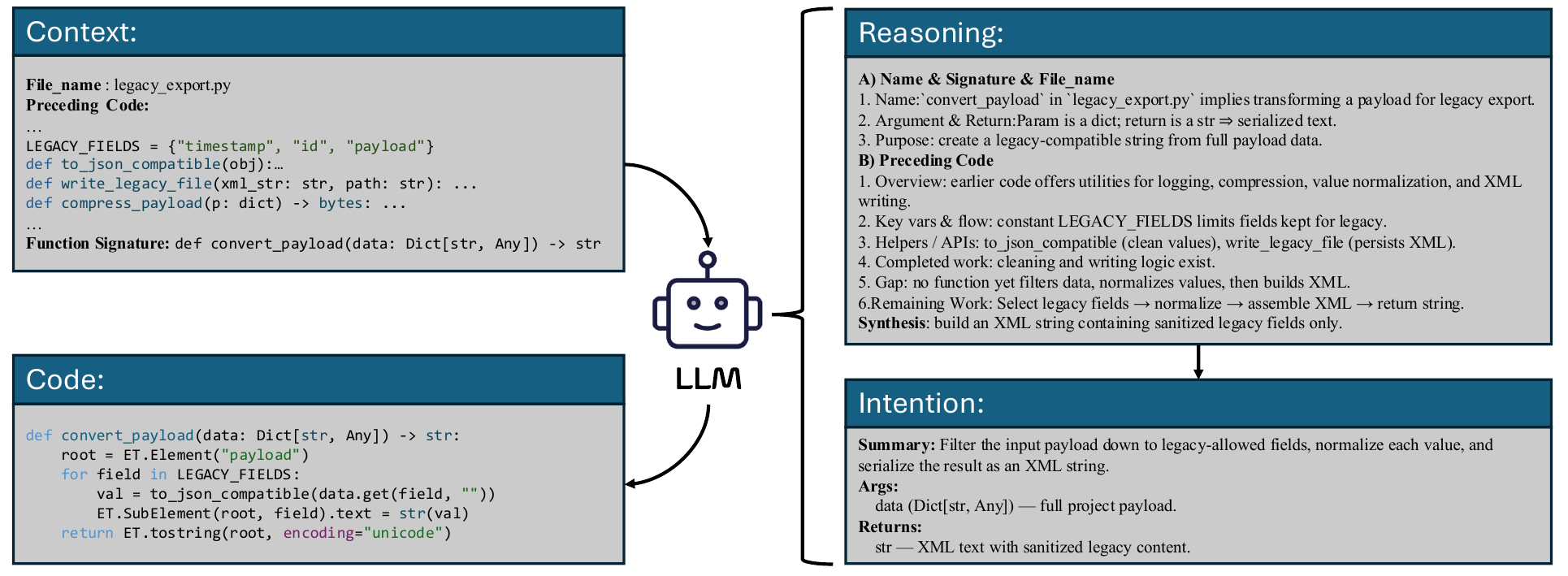}
    \caption{
    Illustration of reasoning process
    }
    \label{fig:reasoning}
    \vspace{-8pt}
\end{figure}

Accurately inferring a function’s intention requires a comprehensive understanding of its surrounding context. This context contains both lexical and semantic cues that provide valuable insights into the expected functionality. Lexical signals such as the file name, function name, and argument names often carry descriptive information. For instance, in a file named \textit{transaction\_utils.py}, a function called \textit{validate\_signature} likely pertains to verifying cryptographic signatures as part of a broader transaction processing pipeline. These names help convey both the high-level purpose of the file and the specific role of the current function. In addition to lexical cues, the preceding code offers semantic context through implemented logic, partially completed workflows, and already resolved sub-tasks. Capturing these elements is essential for understanding what functionality remains unimplemented and what the current function is expected to achieve.

Motivated by this, we propose a structured chain-of-thought template to guide the model in analyzing and synthesizing contextual information. Specifically, each example presents the model with the file name, function signature (including the function name and arguments), and the preceding code. The prompt template is organized into three main components, each comprising several sub-steps.

\paragraph{\textbf{A.} Lexical Cues from Naming}
The prompt enables the model to start by interpreting lexical cues, forming an initial hypothesis about the function's role.
\textbf{(1)} It first prompts the LLMs to analyze the function name and file name to infer the likely action and target domain.
\textbf{(2)} Next, it prompts the LLMs to examine the function's arguments to deduce the roles and expected types of each parameter.
\textbf{(3)} Finally, based on the combined lexical signals, it prompts the model to synthesize a high-level estimate of the function's intended purpose.

\paragraph{\textbf{B.} Semantic Cues from Preceding Code}
Next, the prompt is to enable the model to process the surrounding code in order to extract deeper semantic signals.
\textbf{(1)} It begins by summarizing the overall behavior of the preceding code.
\textbf{(2)} It then identifies key variables, control structures, and helper functions, along with their respective roles.
\textbf{(3)} Finally, the model determines which sub-tasks have already been completed and highlights the remaining functional gap to be addressed.

\paragraph{\textbf{C.} Inferring Intention}
Drawing on insights from Parts A and B, the model is prompted to infer the tasks necessary to complete the target function.
\textbf{(1)} It enumerates the missing logic or functionality the function should implement, such as validation, transformation, or integration with existing components.
\textbf{(2)} If applicable, it also reasons about required control structures, including loops, conditionals, or error handling mechanisms.
\textbf{(3)} The reasoning concludes with a synthesized summary of the function’s intended behavior.

Figure~\ref{fig:reasoning} presents an example of this multi-step reasoning process. It demonstrates the model input, the structured reasoning trace, the inferred intention (expressed as a docstring), and the final generated code. 
In this example, the model first identifies lexical cues such as the function name convert\_payload and the file name legacy\_export.py, which suggest the function's role in preparing data for legacy export. It then analyzes preceding code, noting that helper functions for value normalization and XML writing are already present. By synthesizing these insights, the model determines that the missing functionality involves filtering legacy-specific fields and assembling them into an XML string. The reasoning concludes with a clear and concise docstring that captures this intention, which guides the generation of a well-aligned implementation.

\subsection{Incentivizing LLMs to Infer Intent}
However, prompting is not ideal in our setting as it heavily relies on the reasoning capabilities of LLMs, which are often proprietary. For weaker models, prompting presents three issues: (1) Function-level completion typically involves long code contexts that occupy most of the available input window, leaving insufficient space for demonstrations. (2) Additionally, adding lengthy demonstrations significantly increases prompt size, reducing inference efficiency in real-world applications. (3) Finally, prompting does not reliably enforce structured reasoning behavior.

\begin{figure*}[t]
    \centering
    \includegraphics[width=\textwidth]{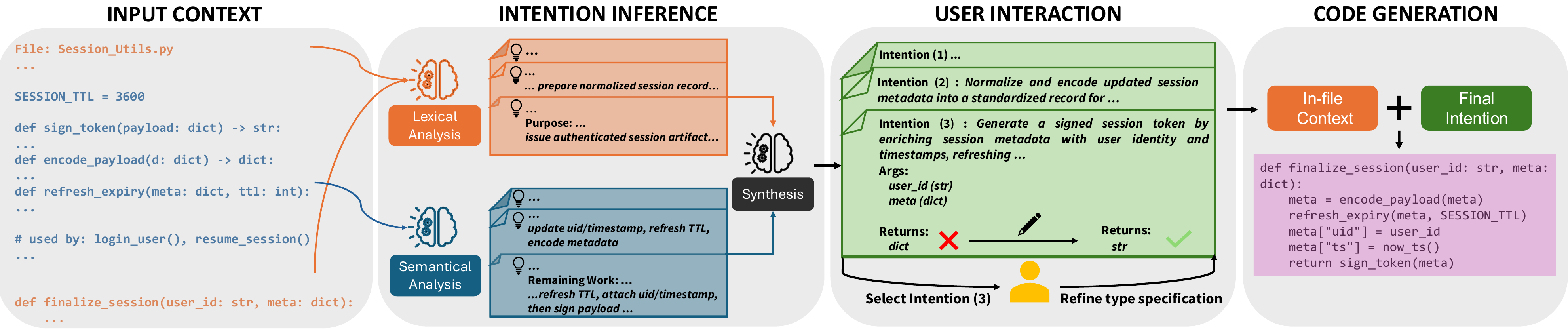}
    \caption{
    Three-stage Completion Framework
    }
    \label{fig:framework}
    \vspace{-8pt}
\end{figure*}

To overcome these limitations, we aim to incentivize the model's intention reasoning capability through a fine-tuning-based approach.
To construct the fine-tuning dataset, we start with a small set of manually annotated examples and use GPT-4o, employing these examples as demonstrations to scale up the dataset.
This yields a training dataset of function-level completion instances, each labeled with intermediate reasoning traces and corresponding docstrings. We then fine-tune code LLMs on this dataset to support accurate and context-aware intention inference.
The following subsections detail our reasoning framework and training methodology.

\textbf{Dataset Construction.}
\label{sec:data}
Specifically, we construct the training data in a four-step pipeline: data sampling, manual annotation, in-context generation, and training data formatting.


\paragraph{\textbf{A.} Data Sampling.}  
To construct training examples, we begin by sampling source files \( C \) from a diverse set of open-source repositories. For each file, we select a function whose body \( f_{\text{body}} \in C \) is used as the completion target. The corresponding context includes the function name, its preceding code, and the function signature, denoted as \( (C_{\text{name}}, C_{\text{pre}}, f_{\text{sig}}) \in C \).


\paragraph{\textbf{B} Manual Annotation.}  To obtain high-quality seed examples, we manually annotate a small set of instances. Given the context \( (C_{\text{name}}, C_{\text{pre}}, f_{\text{sig}}) \) and the target implementation \( f_{\text{body}} \), annotators create:  
(1) a structured reasoning trace \( r \), following the template in Section~\ref{sec:cot}, with each sub-step limited to fewer than 20 words for clarity; and  
(2) a docstring \( d \), formatted according to standard conventions. As illustrated in Figure \ref{fig:reasoning}, each docstring begins with a one-sentence summary of the function’s goal, followed by 1–3 lines describing key operations, along with short descriptions of each argument and the return value.  
This yields a fully annotated seed instance:  
\[
x = (C_{\text{name}}, C_{\text{pre}}, f_{\text{sig}}, r, d, f_{\text{body}})
\]


\paragraph{\textbf{C.} Large-Scale Annotation via In-Context Generation.}
Manually annotating the entire training dataset with structured reasoning and docstrings is prohibitively expensive. To enable scalable supervision, we leverage a commercial code LLM \( \mathcal{F} \) (GPT-4o) to generate annotations based on a few high-quality examples. Specifically, given three manually annotated seed examples \( D = \{x_1, x_2\} \) as in-context demonstrations and access to the ground-truth function implementation, the model generates a reasoning trace and corresponding docstring for a new instance as follows:
\[
(r, d) = \mathcal{F}(C_{\text{name}}, C_{\text{pre}}, f_{\text{sig}}, f_{\text{body}} \mid D)
\]
This process allows us to expand the dataset efficiently while maintaining alignment with the desired annotation format, after which the generated annotations are filtered through structural and semantic validation stages.


\paragraph{\textbf{D.} Data Formatting}  
To prepare data for fine-tuning, we convert each annotated instance  
\(
x = (C_{\text{name}}, C_{\text{pre}}, f_{\text{sig}}, r, d, f_{\text{body}})
\)  
into a standardized string using a formatting function \(\mathcal{V}\). This function organizes the instance into clearly segmented parts. For the reasoning trace \(r\), docstring \(d\), and function body \(f_{\text{body}}\), we insert special tokens \(T\) to delineate each segment. An example of \(\mathcal{V}(x, T)\) is shown below:

\begin{tcolorbox}[colback=gray!5!white, colframe=gray!30!black, boxrule=0.3pt, arc=2pt, left=4pt, right=4pt, top=2pt, bottom=2pt]
\small
\textit{"file name"}: \texttt{C\_name},\\
\textit{"preceding code"}: \texttt{C\_pre},\\
\textit{"function name \& signature"}: \texttt{f\_sig}\\
\textcolor{orange}{\texttt{<reasoning>}} \texttt{r} \textcolor{orange}{\texttt{</reasoning>}}\\
\textcolor{orange}{\texttt{<docstring>}} \texttt{d} \textcolor{orange}{\texttt{</docstring>}}\\
\textcolor{orange}{\texttt{<code>}} \texttt{f\_body} \textcolor{orange}{\texttt{</code>}}
\end{tcolorbox}

\noindent
Here, the \textcolor{orange}{\texttt{<...>}} tokens are pre-defined special tokens used to segment different parts of the sequence. These tokens help the model distinguish between reasoning, documentation, and code generation phases, while also improving structure adherence and downstream interpretability.

\textbf{Training.}
Given the formatted dataset, we fine-tune the model using supervised learning with standard next-token prediction under teacher forcing. Specifically, we minimize the cross-entropy loss over the target sequence, including both the segmentation tokens and content tokens. Formally, the loss for an instance is defined as:

\[
\mathcal{L} = -\sum_{t \in \mathcal{T}} \log P_M\bigl(y_t \mid \mathcal{V}(C_{\text{name}}, C_{\text{pre}}, f_{\text{sig}}),\, y_{<t}\bigr),
\]
where \( M \) represents the LLM being trained. The target sequence \( y_t \) corresponds to the verbalized output \( \mathcal{V}(d, r, f_{\text{body}}, T) \), starting from the first segmentation token (\texttt{<reasoning>}) to the end of the sequence. We exclude tokens preceding this point from the loss computation, as they are treated solely as input context. This setup encourages the model to first generate the structured reasoning trace, followed by the docstring, and finally the function implementation.


\subsection{Three-Stage Intent-Interactive Framework}

While our reasoning-based inference method enhances the model’s ability to analyze context and synthesize informed intentions, developers may still have diverse goals for the same code context. Consequently, even context-aware models might generate intentions that do not fully align with the user’s actual expectations. To address this, we introduce a lightweight Three-Stage Intent-Interactive Framework, enabling users to guide the model toward their intended functionality. This approach provides developers with a practical tool to influence the completion process when necessary.
The full workflow is organized into a three-stage framework: (1) intention inference, (2) user interaction, and (3) function body generation. An overview of this process is shown in Figure~\ref{fig:framework}. The following subsections describe each stage in detail.

\textbf{Intention Inference.}
At inference time, we use the same data format as in training. Given an input sequence comprising the file name, preceding code, and function signature—concatenated with the initial segmentation token \texttt{<reasoning>}—we represent the input as  
\[
x_{\text{input}} = \mathcal{V}(C_{\text{name}}, C_{\text{pre}}, f_{\text{sig}}, T_{\langle\texttt{reasoning}\rangle}).
\]
The model then autoregressively generates a structured reasoning trace \(r\), followed by the docstring \(d\), and terminates generation upon emitting the closing token \texttt{</docstring>}. To support multiple plausible interpretations, we sample the top-\(k\) candidate docstrings ranked by model confidence. These serve as alternative intention hypotheses for downstream user selection.

\textbf{User Selection/Editing.}
Instead of requiring users to write detailed specifications or full documentation, this stage supports minimal-effort refinement of the model’s inferred intention. Specifically, we provide two modes of user feedback:  
\textbf{(1) Selection.} The user reviews the top-\(k\) candidate docstrings generated during intention inference and selects the one that best aligns with their intended functionality.  
\textbf{(2) Word-level Editing.} The selected docstring can be further refined through minor edits, such as correcting return and argument types, or modifying specific functional keywords.

\textbf{Code Generation.}
Given the completion context and the user-adjusted docstring \( \hat{d} \), we adopt the same verbalization strategy described in Section~\ref{sec:data}. Specifically, we apply the formatting function \( \mathcal{V} \) to construct the input string, appending the segmentation token \texttt{<code>} to indicate the start of code generation. The model then produces the completed function body \( \hat{f}_{\text{body}} \) using greedy decoding:
\[
\hat{f}_{\text{body}} = M(\mathcal{V}(C_{\text{name}}, C_{\text{pre}}, f_{\text{sig}}, \hat{d}, T_{<\texttt{code}>})).
\]
This process enables the model to generate a complete implementation that is both consistent with the refined intention and coherent with the surrounding context.

\section{Implementation}
In this section, we present our implementation details.
Specifically, we describe the full setup for training the model, including data construction, fine-tuning, and inference. We also outline the settings used in our evaluation, covering benchmarks, baselines, evaluation metrics, as well as our simulation of the user interaction stage.
\subsection{Training Dataset}
To build a high-quality dataset for training, we first crawled 400 permissively licensed Python repositories created before 2024, selecting the top-ranked projects by GitHub star count to prioritize mature and widely used codebases. From these repositories, we parsed and extracted all top-level function definitions using Python’s AST. We then applied a series of filtering heuristics to retain only suitable candidates for function-level code completion.

Specifically, we retained functions that satisfy the following criteria:
(1) function length smaller than 100 lines, to avoid overly large implementations;
(2) non-trivial function names (excluding common dunder methods or test cases);
(3) syntactically valid code with no placeholder bodies (e.g., \texttt{pass}, \texttt{...});
(4) moderate cyclomatic complexity ($\leq 25$);
(5) sufficiently informative preceding context (20–800 lines); and
(6) exclusion of auto-generated files and functions containing sensitive keywords such as \texttt{password} or \texttt{token}.
After applying these filtering criteria, we obtained a pool of high-quality function-level candidates.
Moreover, to ensure topic diversity, we categorized repositories based on their associated GitHub topics and performed stratified sampling across topic buckets~\cite{singh1996stratified}. From this pool, we sampled 43K function instances as targets for function-level completion, covering a broad range of real-world domains.



Following prior work that constructs large-scale training data by combining expert-annotated seed examples with LLM-based annotation expansion~\cite{wang2023self-instruct,luo2023wizardcoder,chen2024reinstruct,ahmad2025opencodeinstruct}, we then randomly selected 50 functions spanning different topics for manual annotation. Each instance was annotated by a pool of five experts with software engineering and AI experience, following the structured reasoning and intention template defined in Section~\ref{sec:cot}. Annotators were provided with the function body and its surrounding context, and were instructed to reconstruct the underlying intent implied by the implementation rather than paraphrasing low-level code details. To reduce individual bias, the seed annotations underwent cross-review among annotators, improving consistency and accuracy while mitigating annotator-specific stylistic variation, resulting in a small set of high-confidence examples.
These expert-labeled examples were subsequently used as few-shot demonstrations to bootstrap large-scale automatic annotation using GPT-4o. For each generation request, two demonstrations were randomly sampled from the annotated seed set to mitigate demonstration bias. During annotation, the model was given access to the full context, including the ground-truth function body, to ensure that the generated reasoning traces and intentions were semantically grounded in the actual implementation. 

All automatically generated annotations were validated before inclusion. We first applied rule-based parsing to ensure structural correctness of reasoning traces and the presence of required intention fields. We then used an independent LLM (o3-mini) to perform semantic alignment checks, filtering out cases where the inferred intention conflicts with the ground-truth implementation. We additionally randomly sampled 100 instances for human audit to verify overall annotation quality and confirm the absence of systematic issues, such as reasoning hallucination or consistent intent–implementation misalignment. This process resulted in a final dataset of 40K function-level examples annotated with structured reasoning traces and docstrings for training our reasoning-enhanced model.


\subsection{Fine-tuning}
We fine-tune models from two different code LLM families on our constructed dataset, covering a range of model sizes from 7B to 16B. Specifically, we train CodeLlama-7B and CodeLlama-13B, as well as DeepseekCoder-V2-Lite (16B)~\cite{roziere2023codellama,zhu2024deepseekcoder}. Due to resource constraints, we apply full-parameter fine-tuning for the 7B model, while employing parameter-efficient fine-tuning using LoRA~\cite{hu2022lora} for the 13B and 16B models.

We fine-tune 7B models using the AdamW~\cite{kingma2014adam} optimizer with $\beta_1 = 0.9$, $\beta_2 = 0.95$, a learning rate of $1\mathrm{e}{-5}$, and a cosine learning rate schedule with 5\% warmup over 3 epochs. Each training instance is truncated to 4096 tokens by discarding the earliest context tokens. We employ a token-based batch packing strategy with a maximum of 4096 tokens per batch.
For parameter-efficient fine-tuning. We use a learning rate of $1\mathrm{e}{-4}$ and set LoRA-specific hyperparameters as follows: rank $r=32$, $\alpha=64$, and dropout $=0.05$. Following common practice, we target attention and feedforward projections and allow updates to both the input embeddings. Token packing and training configurations remain the same as in the full fine-tuning setup. All training runs are conducted on 4 NVIDIA A100 GPUs (80GB each) using the FlashAttention framework for efficient memory and compute utilization.

\subsection{Inference}
For inference, we append \texttt{<reasoning>} to the verbalized context to trigger generation of the reasoning trace and docstring, ending at \texttt{</docstring>}. We use top-$p$ sampling ($p=0.95$, temperature $0.4$) to generate $k{=}3$ docstring candidates. For code generation, we append \texttt{<code>} to the selected or edited docstring and apply greedy decoding (temperature $0.2$). All inference is conducted using the VLLM framework for efficient batched decoding.

\subsection{User Interaction Simulation}

Since our completion framework incorporates a human-in-the-loop component, we simulate user interaction for large-scale evaluation. For the selection step, we use the \texttt{all-MiniLM-L6-v2} model to embed both model-generated and oracle docstrings, and select the candidate with the highest cosine similarity to the oracle as the one most semantically aligned with user intent. For the editing step, we simulate lightweight user corrections by replacing 1–3 words of argument or return identifiers in the selected docstring based on the oracle, while keeping all natural language descriptions unchanged.

\section{Evaluation}
We aim to answer the following research questions:
\label{sec:eval}
\begin{itemize}
  \item \textbf{RQ1 (Effectiveness of Reasoning):} Does our reasoning-enhanced fine-tuned model outperform baseline LLMs in function-level code completion?

  \item \textbf{RQ2 (Effectiveness of User Interaction):} Could the lightweight intent-interactive framework better align the generated intentions with the user’s actual goals, thereby improving function completion performance?

  \item \textbf{RQ3 (Ablation of Reasoning Components):} What is the impact of reasoning and its components during fine-tuning on intention inference and function completion performance?

  \item \textbf{RQ4 (Efficiency--Effectiveness Trade-off):}
  How does reasoning-based intention inference affect the inference efficiency of function-level code completion models, and what is the efficiency–effectiveness trade-off?

  \item \textbf{RQ5 (Cross-Language Generalization):} To what extent does our approach generalize to a different programming language (Java)?

  \item \textbf{RQ6 (Alternative Deployment Setting):} Can our fine-tuned model serve as a plug-and-play intention inference module to improve the code completion of other models?
\end{itemize}

\subsection{Evaluation Setup}
\subsubsection{Benchmarks}
We evaluate our models on two repository-level function generation benchmarks: DevEval~\cite{li2024deveval} and ComplexCodeEval~\cite{feng2024complexcodeeval}.
\textbf{DevEval} contains 1,825 Python functions with accompanying unit tests, collected from 115 open-source repositories. Each instance includes rich annotations, but we use only the file name, preceding code, and function signature as input, while the annotated function body and docstring serve as oracle references. To ensure the reliability of execution-based evaluation, we filter out instances that do not pass their associated unit tests in our local environment, resulting in 1,211 valid examples. \textbf{ComplexCodeEval} consists of 7,184 Python functions drawn from GitHub projects that use third-party libraries. To avoid potential overlap with the pre-training data of the tested models, we select a subset of 773 instances whose source files were created after 2024 for evaluation. We further confirm that there is no overlap between our training data and the instances used in either benchmark.

\subsubsection{Models\&Baselines}
We evaluate the same model families as used in training: CodeLlama-7B, CodeLlama-13B, and 
DeepSeek-Coder-V2-Lite-Instruct(16B). As baselines, we use the original models without reasoning supervision. All models take identical contexts as input and share the same decoding setup for fair comparison. For baseline generation, we consider two settings: (1) direct completion (direct), where the model generates the function body; and (2) explicit intention-first (intent), where the model is prompted to first generate a docstring representing the intended functionality before completing the corresponding code.

\subsubsection{Evaluation Metrics}
We use reference-based metrics, including CodeBLEU and EditSimilarity to assess functional and syntactic similarity between generated and reference implementations. Additionally, for DevEval, we report pass@1 using the provided unit tests and execution environment to measure functional correctness.

\subsection{RQ1: The Effectiveness of Inferring Intention}
We evaluate our reasoning-enhanced method on DevEval and ComplexCodeEval, comparing it against two baselines: (1) ``direct'' generation, where the model completes the function body directly from context, and (2) ``intent''-first generation, where the model is first prompted to predict a docstring before generating the code. As shown in the left half of Table~\ref{tab:rq1}, simply prompting the model to generate a docstring (intent) brings only minor improvements over the direct variant—for example, for CodeLlama-13B on DevEval, Pass@1 increases by only 0.3\%, and CodeBLEU even drops slightly. This suggests that surface-level intent modeling lacks sufficient guidance. In contrast, our reasoning-enhanced model yields substantial gains. On DevEval, CodeLlama-13B’s Pass@1 improves from 18.99\% (direct) to 27.81\%, while similar relative gains are observed across CodeLlama-7B and DeepSeek-Coder-V2-Lite(16B). Averaged across all models, our method improves Pass@1 by over 7.5\% (40\% relative). CodeBLEU and EditSim also consistently improve, with gains of at least 5.5\% and often higher. On ComplexCodeEval, our model achieves similar gains in BLEU and EditSim across all models, further confirming the benefit of reasoning-based intention modeling.

\begin{table}[t]
\caption{
Function-level code completion results on DevEval and ComplexCodeEval. 
\textbf{CL} = CodeLlama, \textbf{DS} = DeepSeekCoder.  
``C-BLEU'' and ``ES'' denote CodeBLEU and Edit Similarity, respectively.  
``Direct'' refers to vanilla next-token generation; ``Intent'' includes an intermediate docstring generation step; and ``Reason'' uses our fine-tuned reasoning-enhanced model.
}
\label{tab:rq1}
\vspace{-8pt}
\centering
\setlength{\tabcolsep}{4pt}
\renewcommand{\arraystretch}{1.05}
\resizebox{\linewidth}{!}{
\begin{tabular}{ll|ccc|cc||ccc|cc}
\toprule
\multirow{2}{*}{Model} & \multirow{2}{*}{Variant} 
& \multicolumn{5}{c||}{\textbf{In-file Setting}} 
& \multicolumn{5}{c}{\textbf{RAG Setting}} \\
\cmidrule{3-12}
& & \multicolumn{3}{c|}{\textbf{DevEval}} & \multicolumn{2}{c||}{\textbf{ComplexCodeEval}}
& \multicolumn{3}{c|}{\textbf{DevEval}} & \multicolumn{2}{c}{\textbf{ComplexCodeEval}} \\
\cmidrule{3-12}
& & C-BLEU & ES & P@1 & C-BLEU & ES & C-BLEU & ES & P@1 & C-BLEU & ES \\
\midrule
\multirow{3}{*}{CL-7B} 
& direct    & 38.48 & 39.93 & 17.51 & 29.15 & 32.86 & 40.38 & 43.17 & 18.00 & 33.06 & 34.17 \\
& intent    & 38.69 & 40.47 & 17.56 & 26.58 & 30.31 & 41.29 & 43.88 & 19.32 & 31.72 & 32.50 \\
& reason (ours) & \textbf{46.68} & \textbf{48.61} & \textbf{26.34} & \textbf{32.83} & \textbf{36.97} 
& \textbf{48.20} & \textbf{50.98} & \textbf{27.83} & \textbf{38.42} & \textbf{40.81} \\
\midrule
\multirow{3}{*}{CL-13B} 
& direct    & 41.36 & 43.60 & 18.99 & 31.11 & 34.71 & 42.10 & 45.26 & 20.48 & 34.95 & 36.01 \\
& intent    & 39.61 & 42.79 & 19.32 & 29.61 & 33.61 & 42.47 & 44.02 & 20.56 & 33.87 & 34.80 \\
& reason (ours) & \textbf{47.45} & \textbf{48.18} & \textbf{27.81} & \textbf{35.60} & \textbf{38.05} 
& \textbf{50.74} & \textbf{52.36} & \textbf{28.74} & \textbf{40.50} & \textbf{41.33} \\
\midrule
\multirow{3}{*}{DS-16B} 
& direct    & 42.05 & 42.80 &21.29  & 32.21 & 35.29 & 43.67 & 44.00 & 21.97 & 36.20 & 37.17 \\
& intent    & 41.43 & 43.96 & 20.97 & 31.35 & 34.12 & 43.52 & 43.81 & 22.21 & 35.79 & 36.83 \\
& reason (ours) & \textbf{47.24} & \textbf{49.55} & \textbf{29.48} & \textbf{37.59} & \textbf{40.40} 
& \textbf{49.73} & \textbf{53.07} & \textbf{29.07} & \textbf{43.37} & \textbf{45.28} \\
\bottomrule
\end{tabular}}

\vspace{-10pt}
\end{table}

Recent work on code completion has increasingly focused on repository-level completion, where the goal is to leverage project context beyond a single file. A prominent line of research adopts retrieval-augmented generation (RAG) to improve completion quality by retrieving relevant code modules from other files in the same repository. These methods typically identify cross-file dependencies based on the preceding code and incorporate them into the input prompt~\cite{zhang2023repocoder,cheng2024draco,wu2024repoformer}. To ensure broader applicability, we additionally evaluate our method under an RAG setting that incorporates cross-file context. In particular, we aim to examine whether our reasoning-based intention inference complements retrieval-based methods and can provide additional improvements when used together.
As shown in the RAG setting columns of Table~\ref{tab:rq1}, all models and generation variants benefit from RAG, with baseline CodeBLEU scores increasing by 1–3 points. More importantly, our model maintains a consistent advantage—for example, on DevEval, reasoning-based CodeLlama-13B improves CodeBLEU from 42.10 (direct) to 50.74, and EditSim from 45.26 to 52.36. These margins are comparable to those in the in-file setting, indicating that our structured reasoning remains effective even when augmented with cross-file dependencies.

\begin{tcolorbox}[highlightbox]
\textbf{Answer for RQ1:} Our reasoning-based model consistently outperforms both direct and intent-based baselines in function-level code completion across two benchmarks. It improves Pass@1 by over 8.5\% on average and raises CodeBLEU and EditSim by at least 4\%. These performance gains hold across different model sizes and settings, including both in-file and RAG-enhanced scenarios, demonstrating the robustness and effectiveness of our method.
\end{tcolorbox}

\definecolor{oracleyellow}{RGB}{255, 245, 190}
\begin{table}[t]
\caption{Evaluation of the intent-interactive framework. ``Sim'' denotes the cosine similarity.}
\label{tab:rq2}
\vspace{-8pt}
\centering
\small
\setlength{\tabcolsep}{8pt}
\renewcommand{\arraystretch}{1.05}
\resizebox{0.95\linewidth}{!}{
\begin{tabular}{ll|cccc|ccc}
\toprule
\multirow{2}{*}{\textbf{Model}} & \multirow{2}{*}{\textbf{Variant}} 
& \multicolumn{4}{c|}{\textbf{DevEval}} 
& \multicolumn{3}{c}{\textbf{Complex}} \\
\cmidrule{3-9}
& & C-BLEU & ES & P@1 & Sim & C-BLEU & ES & Sim \\
\midrule

\multirow{6}{*}{CL-7B}
& oracle      & 51.81 & 51.04 & 32.29 & - & 35.48 & 38.87 & - \\\hdashline
& org      & 38.48 & 39.93 & 17.51 & 63.18 & 29.15 & 32.86 & 66.40 \\
& reason      & 46.68 & 48.61 & 26.34 & 74.09 & 32.83 & 36.97 & 78.80 \\
& +select       & 46.67 & 48.85 & 27.75 & 76.53 & 33.96 & 37.33 & 79.52 \\
& +edit     & 46.02 & 48.96 & 27.83 & 74.38 & 33.53 & 37.37 & 79.18 \\
& +both       & \textbf{47.27} & \textbf{49.80} & \textbf{29.15} & \textbf{76.93} & \textbf{34.01} & \textbf{38.15} & \textbf{80.10} \\
\midrule

\multirow{6}{*}{CL-13B}
& oracle      & 53.19 & 54.71 & 35.18 & - & 38.49 & 42.95 & - \\\hdashline
& org      & 41.36 & 43.60 & 18.09 & 65.81 & 31.11 & 34.71 & 67.73 \\
& reason      & 47.45 & 48.18 & 27.81 & 73.85 & 35.60 & 38.05 & 77.96 \\
& +select       & 49.14 & 50.57 & 28.65 & 75.52 & \textbf{37.41} & \textbf{40.20} & 79.92 \\
& +edit     & 48.34& 49.20 & 29.56 & 74.15 & 35.47 & 37.31 & 78.46  \\
& +both       & \textbf{50.62} & \textbf{51.80} & \textbf{30.72} & \textbf{76.33} & 35.79 & 39.67 & \textbf{80.10} \\

\midrule

\multirow{6}{*}{DS-16B}
& oracle      & 53.55 & 52.51 & 34.27 & - & 39.82 & 42.63 & - \\\hdashline
& org      & 42.05 & 42.80 & 21.29  & 68.73 & 32.21 & 35.29 & 69.08 \\
& reason      & 47.24 & 49.55 & 29.48 & 75.06 & 37.59 & 40.40 & 79.97 \\
& +select      & 50.89 & 50.99 & 30.05 & 78.33 & 37.46 & \textbf{42.02} & 81.04 \\
& +edit     & 47.85 & 49.14 & 31.30 & 75.78 & 36.89 & 41.54 & 80.26 \\
& +both       & \textbf{51.49} & \textbf{51.34} & \textbf{31.71} & \textbf{78.47} & \textbf{38.17} & 41.96 & \textbf{81.52} \\
\bottomrule
\end{tabular}}
\vspace{-10pt}
\end{table}

\subsection{RQ2: Intention-Interactive Code Completion}
To evaluate whether the intention-interactive framework can align model-generated intentions more closely with user expectations and improve function-level code completion, we conduct experiments on both types of interaction supported in our framework: selection and lightweight editing. We compare six variants. (1)“Oracle”: the model is provided with the ground-truth intention (i.e., the reference docstring) before generating the code. (2) ``intent'': baseline model with automatically generated intention, (3) ``reason'': our reasoning model without interaction, (4) ``+select'': reasoning model with selection, (5) ``+edit'': reasoning model with editing, and (6) ``+both'': reasoning model with both selection and editing. In addition to the standard completion metrics, we introduce a new evaluation metric: the semantic alignment between the model-generated intention and the oracle intention. Specifically, we compute the cosine similarity between their sentence embeddings using the pre-trained encoder model all-MiniLM-L6-v2. This allows us to assess whether user interaction produces intentions that are semantically closer to what the function is actually designed to do.

The experimental results are shown in Table~\ref{tab:rq2}. Both operations consistently improve completion performance over our base reasoning model across all three model sizes. On the DevEval benchmark, editing brings an average improvement of 1.7\% in Pass@1, while selection contributes approximately 1\%. When combined, they yield a cumulative gain of about 2.6\%, underscoring the value of lightweight human interaction. These improvements are also reflected in the reference-based metrics. Moreover, both operations substantially increase the semantic alignment between generated and oracle intentions, with the combined variant achieving the highest similarity score i.e., up to 76.93 for CodeLlama-7B, 76.33 for CodeLlama-13B, and 78.47 for DeepSeekCoder-V2-Lite. Notably, all reasoning-based variants, from the base model to the human-in-the-loop versions, consistently outperform the intent-based baseline by more than 10\% in similarity. This highlights the role of these techniques in aligning code completions more closely with user intent.

As shown in Table~\ref{tab:rq2}, we additionally include an oracle setting where the model is provided with the ground-truth docstring prior to code generation, serving as an upper bound for completion performance. These rows are highlighted in yellow. Compared to the oracle, intent-based baselines exhibit a substantial performance gap, often exceeding 15\% across all metrics. In contrast, our reasoning-enhanced model, when combined with lightweight user interaction, significantly narrows this gap to under 4\% on average—an acceptable margin. In practice, user-written instructions frequently miss key intent details and rarely achieve oracle-level completeness. Our approach offers a practical alternative, enabling automated function completion with minimal performance loss.

To verify the fidelity of our interaction simulation, we assess whether the simulated selection and editing operations align with human preferences. Editing is correct by construction, as it directly fixes objective intent attributes based on the ground-truth code. For selection, we randomly sample 100 instances and ask five domain experts to choose the most appropriate intention from the same candidate set using the ground-truth implementation as reference. The majority-voted choice is treated as the human-preferred intention and compared with the model’s selection. We observe 81\% agreement between model-selected and human-preferred intentions, indicating that the selection-based simulation closely reflects human judgment. Both interaction mechanisms are lightweight and avoid the need for extensive docstring rewriting.

\definecolor{lightgray}{gray}{0.95}
\begin{table}[t]
\caption{
Ablation study on reasoning strategies during fine-tuning for \textbf{CodeLlama-7B}.
}
\label{tab:rq3}
\vspace{-6pt}
\centering
\small
\setlength{\tabcolsep}{8pt}
\renewcommand{\arraystretch}{1.05}
\resizebox{0.95\linewidth}{!}{%
\begin{tabular}{l|cccc|ccc}
\toprule
& \multicolumn{4}{c|}{\textbf{DevEval}} & \multicolumn{3}{c}{\textbf{Complex}} \\
\textbf{Reasoning Setting} 
& C-BLEU & ES & P@1 & Sim
& C-BLEU & ES & Sim \\
\midrule
no reasoning     & 40.86 & 42.95 & 19.74 & 71.34 & 30.08 & 33.59 & 73.97 \\
lexical-only     & 45.23 & 45.88 & 23.86 & 72.47 & 31.66 & 36.48 & 77.51 \\
semantic-only    & 44.16 & 47.92 & 25.10 & 32.20 & 31.46 & 35.80 &\textbf{79.12} \\
full reasoning  & \textbf{46.68} & \textbf{48.61} & \textbf{26.34} & \textbf{74.09} & \textbf{32.83} & \textbf{36.97} & 78.80  \\
\bottomrule
\end{tabular}}
\vspace{-6pt}
\end{table}

\begin{tcolorbox}[highlightbox]
\textbf{Answer for RQ2:} The lightweight intent-interactive framework, including candidate selection and minimal editing, consistently improves the alignment between generated and oracle intentions. This alignment translates into better code completion performance, narrowing the gap to oracle-level quality to within 3\% across key metrics on average.
\end{tcolorbox}

\subsection{RQ3: Impact of Reasoning Components During Fine-Tuning}
To understand the contribution of different components in our intention inference framework, we conduct an ablation study on the reasoning steps used during fine-tuning. Our approach synthesizes both lexical cues (e.g., function name, signature, and file name) and semantic cues (e.g., preceding code context) to construct a structured intention. To isolate the impact of each, we create four variants of the training data: (1) no reasoning—where the model generates docstring and code directly from context, (2) lexical-only—using only surface-level cues, (3) semantic-only—using only preceding code, and (4) full reasoning—combining both cue types.
We fine-tune CodeLlama-7B on each variant with consistent training and inference settings. As shown in Table~\ref{tab:rq3}, the full reasoning variant achieves the best performance across all metrics. For example, on DevEval, Pass@1 improves from 19.74\% of no reasoning to 26.34\%, and EditSim from 42.95\% to 48.61\%. This highlights that the performance gain stems not just from fine-tuning, but specifically from learning how to reason about function intent. Both the lexical-only and semantic-only settings show notable gains over the no-reasoning baseline. However, neither outperforms the full reasoning model, indicating that lexical and semantic reasoning provide complementary signals necessary for high-quality intention inference and code generation.

\begin{tcolorbox}[highlightbox]
\textbf{Answer for RQ3:} Full reasoning supervision, which combines lexical and semantic cues, consistently leads to better intention alignment and code completion performance. While both components are individually helpful, their combination yields the highest gains, confirming their complementary roles in the intention inference process.
\end{tcolorbox}

\subsection{RQ4: Inference Efficiency of Our Reasoning-based Function Completion} 

To assess the computational efficiency of our proposed approach, we evaluate its inference overhead relative to baseline strategies. Since our method introduces an additional reasoning stage before generating the docstring and final code, it inevitably incurs extra latency. While optional human interaction steps such as editing and selection are valuable, they are difficult to quantify uniformly and are therefore excluded from this analysis. We focus on the non-interactive setting, where the model performs reasoning and code generation in a single pass without user input.

We report two standard metrics: latency, measured as the average time (in seconds) required to complete a function-level prediction, and throughput, measured as the number of functions completed per second. All experiments are conducted using vLLM on an A100-80GB GPU with a batch size set to 1. The average input length is approximately 3,187 tokens. Variants differ in the amount of generated tokens: Direct produces only code, Intent generates both docstring and code, and Reasoning includes reasoning steps, docstring, and code. These differences directly influence latency and throughput.

\begin{table}[t]
\caption{Estimated function-level inference efficiency of CodeLlama-7B under different generation strategies.}
\label{tab:rq4}
\vspace{-8pt}
\centering
\small
\setlength{\tabcolsep}{3pt}
\renewcommand{\arraystretch}{1.05}
\begin{tabular}{l|ccc}
\toprule
\textbf{Variant} & \textbf{Gen\_Tokens} & \textbf{Latency (s/func)} & \textbf{Throughput (func/s)} \\
\midrule
Direct     & 142  & 0.61  & 1.64 \\
Intent     & 264  & 1.05  & 0.95 \\
Reason  & 385  & 1.44  & 0.69 \\
\bottomrule
\end{tabular}
\vspace{-8pt}
\end{table}

As shown in Table~\ref{tab:rq4}, our reasoning-enhanced model introduces additional latency due to the increased generation length. Specifically, it adds about 0.4 seconds per function compared to the intent variant. However, the overall latency remains under 1.5 seconds per function, and throughput stays around 0.7 functions per second, indicating that the method remains practical for real-world usage. Given the substantial improvements in completion quality, this trade-off is acceptable.

\begin{wrapfigure}{r}{0.48\linewidth}
\vspace{-8pt}
\centering
\includegraphics[width=\linewidth]{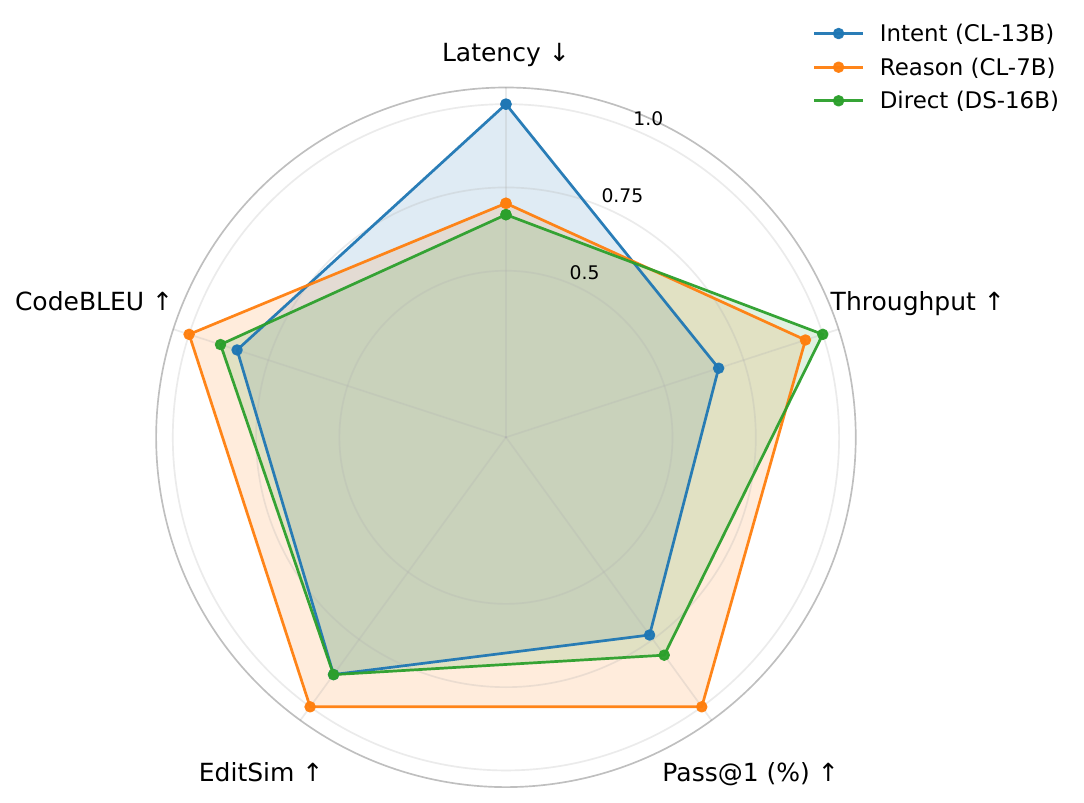}
\vspace{-8pt}
\caption{
Trade-off between efficiency and completion quality.
}
\label{fig:cost_tradeoff}
\vspace{-10pt}
\end{wrapfigure}

To further examine the efficiency–effectiveness trade-off, we consider three generation strategies with different token budgets and pair them with models of increasing scale (7B, 13B, and 16B) to form cost-aware configurations. Specifically, we combine larger models with more efficient generation strategies and smaller models with more expressive reasoning-based strategies, enabling a fair comparison under comparable inference budgets.
Under this setting, Figure~\ref{fig:cost_tradeoff} visualizes the normalized trade-off across different model–strategy pairs.

We first compare our reasoning-enhanced CodeLlama-7B model with the direct generation using DeepSeekCoder(16B), as these two configurations exhibit similar efficiency in terms of latency and throughput. Despite having comparable inference cost, the reasoning-based 7B model achieves substantially better completion quality across all three metrics. We further include a prompt-based intention generation baseline using CodeLlama-13B model. Although this configuration incurs significantly higher latency and lower throughput, it still underperforms the reasoning-based 7B model in completion quality by a big margin, highlighting the advantage of reasoning-based alignment in improving generation effectiveness without sacrificing cost.

\begin{tcolorbox}[highlightbox]
\textbf{Answer for RQ4:} Our reasoning-based intent inference method incurs minimal latency overhead and maintains strong throughput. Compared to stronger baselines under equivalent inference cost, it delivers significantly better completion quality, validating its practical efficiency–effectiveness trade-off.
\end{tcolorbox}

\subsection{RQ5: Generalization to JAVA}
Although our training data consists exclusively of Python functions, we hypothesize that the core capabilities learned by our model, namely understanding code context and inferring high-level developer intention, are not inherently tied to a specific programming language. While programming languages differ substantially in syntax and idioms, the reasoning required to interpret surrounding context and recover intended functionality often follows similar high-level principles. Based on this observation, we investigate whether our reasoning-enhanced model exhibits generalization ability when applied to a different programming language without additional fine-tuning.

\begin{table}[t]
\caption{Performance on the Java subset of ComplexCodeEval. 
$\Delta$ denotes the absolute improvement over \emph{Direct} generation.}
\label{tab:rq5-java}
\vspace{-6pt}
\centering
\small
\setlength{\tabcolsep}{8pt}
\renewcommand{\arraystretch}{1.15}
\begin{tabular}{l|cc|cc||cc|cc}
\toprule
\multirow{2}{*}{Method} 
& \multicolumn{4}{c||}{\textbf{CodeLLaMA-13B}} 
& \multicolumn{4}{c}{\textbf{DeepSeek-Coder-V2-Lite(16B)}} \\
\cmidrule(lr){2-5} \cmidrule(lr){6-9}
& \multicolumn{2}{c|}{ES} & \multicolumn{2}{c||}{CodeBLEU}
& \multicolumn{2}{c|}{ES} & \multicolumn{2}{c}{CodeBLEU} \\
\cmidrule(lr){2-3} \cmidrule(lr){4-5} \cmidrule(lr){6-7} \cmidrule(lr){8-9}
& Score & $\Delta$ & Score & $\Delta$
& Score & $\Delta$ & Score & $\Delta$ \\
\midrule
Direct 
& 20.12 & -- & 17.84 & -- 
& 27.33 & -- & 23.15 & -- \\

Direct-Intent 
& 21.46 & +1.34 & 16.90 & -0.94 
& 28.18 & +0.75 & 24.06 & +1.91 \\

Reasoning (Ours) 
& \textbf{23.67} & \textbf{+3.55} & \textbf{22.59} & \textbf{+4.75}
& \textbf{29.80} & \textbf{+2.47} & \textbf{25.46} & \textbf{+2.31} \\

\bottomrule
\end{tabular}
\vspace{-6pt}
\end{table}

To this end, we extend our evaluation to the Java subset of ComplexCodeEval and conduct experiments on two representative model families, namely CodeLLaMA-13B and DeepSeek-Coder-V2-Lite. To ensure robust generalization, the model used for this evaluation was trained on Python data with a conservative learning rate of 1e-6, and then applied directly to Java instances without any language-specific fine-tuning. We consider three settings, including direct generation, direct-intent generation, and our reasoning-based approach. For both the direct-intent and reasoning-based settings, the prompt includes a single example illustrating Java-style docstring conventions. Following prior experiments, we evaluate completion quality using Edit Similarity (ES) and CodeBLEU, which measure textual similarity and structural correspondence, respectively.

The quantitative results are summarized in Table \ref{tab:rq5-java}. Overall, the reasoning-based approach achieves higher ES and CodeBLEU scores than both direct generation and direct-intent baselines in the Java setting. Although the absolute performance is lower than that observed in Python, which is expected due to language-specific differences and distribution shifts, the relative improvements remain consistent across settings. These results indicate that the reasoning and intention inference mechanisms learned from Python data can transfer meaningfully to another programming language, despite substantial syntactic and idiomatic differences.

\begin{tcolorbox}[highlightbox]
\textbf{Answer for RQ5:} 
Although trained solely on Python data, our model exhibits consistent performance gains when transferred to Java. While the absolute improvements are smaller than those observed in Python, the reasoning-based approach remains effective across different model families and evaluation metrics, indicating that the learned intention inference mechanism generalizes beyond a single programming language.
\end{tcolorbox}

\subsection{RQ6: Plug-and-play Intention Inference Module}

To explore the generalizability and practical utility of our intention inference method, we evaluate a plug-and-play setting where a smaller fine-tuned model (e.g., CodeLlama-7B) is used to generate an intent (i.e., a docstring), which is then passed to a stronger LLM for function completion. This setup investigates whether reasoning performed by a lightweight model can benefit large proprietary models such as GPT-4o, o3-mini~\cite{hurst2024gpt}, and DeepSeek-V3~\cite{liu2024deepseekv3}, enabling a cost-effective and modular generation pipeline. We compare three generation variants for each large model: (1) Direct, where the model generates code from context without an explicit intent; (2) 2-shots, where the model performs in-context reasoning using our prompting template guided by two demonstrations; and (3) Plug-in, where the small model infers the intent, and the large model generates code based on it. Results on DevEval are shown in Table~\ref{tab:rq5}.

Across all three models, the plug-in setting consistently outperforms the direct approach in terms of Pass@1. For example, on GPT-4o, Pass@1 increases from 29.56 to 34.66 and EditSim from 45.78 to 48.59. Similar gains are observed on o3-mini and DeepSeek-V3.
Compared to zero-shot prompting, the plug-in strategy achieves better or comparable Pass@1 scores while using a much smaller model for reasoning. On average, plug-in improves Pass@1 by 1.9 points over 2-shots reasoning. This demonstrates that the intentions inferred by our fine-tuned model can generalize across models and serve as effective guidance for a variety of strong LLMs. In particular, DeepSeek-V3 achieves the best overall performance under the plug-in setting across all metrics. These results validate that our model can act as a lightweight and effective intention inference module, enhancing code generation quality without incurring additional reasoning overhead.

\begin{tcolorbox}[highlightbox]
\textbf{Answer for RQ6:} Our fine-tuned model serves as an effective plug-and-play intention inference module. When paired with stronger LLMs, it consistently improves code completion performance and even outperforms zero-shot reasoning in Pass@1, despite using a much smaller model. This demonstrates that our inferred intents generalize well and offer a lightweight, modular solution for enhancing large-scale code generation.
\end{tcolorbox}

\begin{table}[t]
\caption{
Code generation results on \textbf{DevEval} using different intention provisioning strategies on large models.
$\Delta$ denotes the absolute improvement over \emph{Direct}.
}
\label{tab:rq5}
\vspace{-6pt}
\centering
\small
\setlength{\tabcolsep}{8pt}
\renewcommand{\arraystretch}{1.05}
\resizebox{0.85\linewidth}{!}{
\begin{tabular}{llcc|cc|cc}
\toprule
\textbf{Model} & \textbf{Setting} 
& \multicolumn{2}{c|}{\textbf{Pass@1} ↑}
& \multicolumn{2}{c|}{\textbf{EditSim} ↑}
& \multicolumn{2}{c}{\textbf{CodeBLEU} ↑} \\
\cmidrule(lr){3-4} \cmidrule(lr){5-6} \cmidrule(lr){7-8}
& & Score & $\Delta$ & Score & $\Delta$ & Score & $\Delta$ \\
\midrule
\multirow{3}{*}{GPT-4o}
& Direct      & 29.56 & --   & 45.78 & --   & 44.82 & --   \\
& 2-shots   & 33.77 & +4.21 & 47.43 & +1.65 & \textbf{49.40} & \textbf{+4.58} \\
& Plug-in     & \textbf{34.66} & \textbf{+5.10} & \textbf{48.59} & \textbf{+2.81} & 47.74 & +2.92 \\
\midrule
\multirow{3}{*}{o3-mini}
& Direct      & 30.88 & --   & 43.06 & --   & 42.56 & --   \\
& 2-shots   & 33.28 & +2.40 & \textbf{45.78} & \textbf{+2.72} & \textbf{47.14} & \textbf{+4.58} \\
& Plug-in     & \textbf{35.66} & \textbf{+4.78} & 44.67 & +1.61 & 43.69 & +1.13 \\
\midrule
\multirow{3}{*}{DeepSeek-V3}
& Direct      & 32.12 & --   & 46.42 & --   & 47.15 & --   \\
& 2-shots   & 35.92 & +3.80 & 49.74 & +3.32 & 47.26 & +0.11 \\
& Plug-in     & \textbf{36.66} & \textbf{+4.54} & \textbf{50.38} & \textbf{+3.96} & \textbf{48.02} & \textbf{+0.87} \\
\bottomrule
\end{tabular}}
\vspace{-8pt}
\end{table}

\section{Related Work}
\label{sec:related}

\textbf{Code Completion.} Recent work on code completion primarily targets line-level generation, where models predict the next token or statement based on the immediate local context. Studies such as comprehensive evaluation of code completion by LLMs~\cite{jiang2024completionsurvey,husein2025completionreview,izadi2024completionevaluation,nguyen2019function}, CodeFill~\cite{izadi2022codefill}, and aiXcoder~\cite{jiang2024aixcoder} explore techniques for structure-aware prediction, sequence modeling, and efficient model design for accurate and fast code completion. Another line of research focuses on RAG frameworks~\cite{lewis2020rag} for repository-level code completion. These methods aim to retrieve relevant cross-file contexts from the current repository to support more informed generation. Benchmarks such as CrossCodeEval~\cite{ding2023crosscodeeval} and RepoBench~\cite{liurepobench} have been introduced to evaluate this setting, while methods like RepoCoder, GraphCodeRetriever, and DraCo~\cite{liu2024graphcoder,cheng2024draco,zhang2023repocoder,bui2024rambo} explore various retrieval strategies, including static analysis, control/data flow modeling, and embedding similarity. Some alternatives like LongCoder~\cite{guo2023longcoder} instead extend the model’s context window to accommodate longer in-repo code sequences. In contrast, our work focuses on modeling and inferring the underlying function intention, an orthogonal and largely underexplored direction.

\textbf{Reasoning and Intention Extraction in Code-related Tasks.}
Recent studies have demonstrated that LLMs possess strong multi-step reasoning abilities~\cite{wei2022cot}, leading to the development of CoT prompting~\cite{diao2024activecot,zhao2024enhancing}. In the code domain, CoT has been applied to code generation, algorithm synthesis~\cite{li2025scodecot,huang2023codecot,zhu2025uncertaintycot,li2024chainofcode}, and debugging~\cite{yang2024codedebug,yang2024coast}, typically decomposing problems into intermediate steps under well-defined prompts. In contrast, our work targets the inference of implicit intention, a higher-level reasoning objective that precedes code generation and lacks explicit supervision. Meanwhile, in the domain of code repair, recent work has explored extracting user intention to guide the fix process~\cite{guo2025intention,ruan2025specrover}, typically by generating specifications from buggy code and accompanying user instructions. While related in motivation, these methods operate under more informative conditions, as they have access to both the faulty function and explicit human-written cues that directly reveal the intended fix. In contrast, our setting provides only the preceding code, requiring the model to infer intention without direct supervision or visibility into the target code. This makes the task inherently more ambiguous and calls for explicit reasoning to identify the purpose from the indirect context. To support such inference efficiently, we design a lightweight and single-pass reasoning framework with optional user interaction, making it well-suited for practical code completion scenarios.

\section{Threats to Validity}
While our method performs well in inferring function-level intentions from preceding code, it faces several validity threats. First, it relies on a fixed reasoning template, which may not generalize across all function types or coding patterns—e.g., asynchronous utilities or system-level code may require more domain-specific cues. Future work will explore more flexible prompting strategies to better handle diverse coding scenarios. 

Additionally, real-world inputs often include informal comments or partial specifications, which can aid or complicate intent inference. We plan to incorporate such cues into the reasoning process using adaptive prompts that adjust based on available context. Second, our main evaluation is limited to Python. Whether the framework generalizes to other languages with different paradigms or type systems remains unclear. Adapting the method for statically typed or low-level languages will be part of future work.

\section{Conclusion}
\label{sec:conclude}

This paper investigates the role of intention inference in enhancing function-level code completion. 
To better leverage context, we draw inspiration from how humans interpret it to propose a reasoning-based prompting framework that guides LLMs to utilize contextual cues for step-by-step intent inference. 
We then curate a dataset to incentivize LLMs to reason through the preceding code and infer intent via finetuning.
Extensive experiments demonstrate the effectiveness of our approach, showing that explicit reasoning benefits both intention inference and code generation quality. These results underscore the value of integrating intention understanding into code completion workflows.

\bibliographystyle{ACM-Reference-Format}
\bibliography{sample-base}

@inproceedings{zhang2023repocoder,
  title={RepoCoder: Repository-Level Code Completion Through Iterative Retrieval and Generation},
  author={Zhang, Fengji and Chen, Bei and Zhang, Yue and Keung, Jacky and Liu, Jin and Zan, Daoguang and Mao, Yi and Lou, Jian-Guang and Chen, Weizhu},
  booktitle={The 2023 Conference on Empirical Methods in Natural Language Processing}
}

@misc{businessofapps_copilot2025,
  author       = "{Business of Apps}",
  title        = "{Microsoft Copilot Revenue and Usage Statistics (2025)}",
  howpublished = "\url{https://www.businessofapps.com/data/microsoft-copilot-statistics/}",
  note         = "Accessed 2025‑06‑18",
  year         = "2025",
  month        = "May"
}

@article{ding2023crosscodeeval,
  title={Crosscodeeval: A diverse and multilingual benchmark for cross-file code completion},
  author={Ding, Yangruibo and Wang, Zijian and Ahmad, Wasi and Ding, Hantian and Tan, Ming and Jain, Nihal and Ramanathan, Murali Krishna and Nallapati, Ramesh and Bhatia, Parminder and Roth, Dan and others},
  journal={Advances in Neural Information Processing Systems},
  volume={36},
  pages={46701--46723},
  year={2023}
}

@inproceedings{feng2024complexcodeeval,
  title={Complexcodeeval: A benchmark for evaluating large code models on more complex code},
  author={Feng, Jia and Liu, Jiachen and Gao, Cuiyun and Chong, Chun Yong and Wang, Chaozheng and Gao, Shan and Xia, Xin},
  booktitle={Proceedings of the 39th IEEE/ACM International Conference on Automated Software Engineering},
  pages={1895--1906},
  year={2024}
}

@article{li2024deveval,
  title={Deveval: Evaluating code generation in practical software projects},
  author={Li, Jia and Li, Ge and Zhao, Yunfei and Li, Yongmin and Jin, Zhi and Zhu, Hao and Liu, Huanyu and Liu, Kaibo and Wang, Lecheng and Fang, Zheng and others},
  journal={arXiv preprint arXiv:2401.06401},
  year={2024}
}

@article{li2024evocodebench,
  title={Evocodebench: An evolving code generation benchmark aligned with real-world code repositories},
  author={Li, Jia and Li, Ge and Zhang, Xuanming and Dong, Yihong and Jin, Zhi},
  journal={arXiv preprint arXiv:2404.00599},
  year={2024}
}

@inproceedings{guo2023longcoder,
  title={Longcoder: A long-range pre-trained language model for code completion},
  author={Guo, Daya and Xu, Canwen and Duan, Nan and Yin, Jian and McAuley, Julian},
  booktitle={International Conference on Machine Learning},
  pages={12098--12107},
  year={2023},
  organization={PMLR}
}

@inproceedings{liu2020multicompletion,
  title={Multi-task learning based pre-trained language model for code completion},
  author={Liu, Fang and Li, Ge and Zhao, Yunfei and Jin, Zhi},
  booktitle={Proceedings of the 35th IEEE/ACM international conference on automated software engineering},
  pages={473--485},
  year={2020}
}

@inproceedings{izadi2024completionevaluation,
  title={Language models for code completion: A practical evaluation},
  author={Izadi, Maliheh and Katzy, Jonathan and Van Dam, Tim and Otten, Marc and Popescu, Razvan Mihai and Van Deursen, Arie},
  booktitle={Proceedings of the IEEE/ACM 46th International Conference on Software Engineering},
  pages={1--13},
  year={2024}
}

@article{liu2024graphcoder,
  title={Graphcoder: Enhancing repository-level code completion via code context graph-based retrieval and language model},
  author={Liu, Wei and Yu, Ailun and Zan, Daoguang and Shen, Bo and Zhang, Wei and Zhao, Haiyan and Jin, Zhi and Wang, Qianxiang},
  journal={arXiv preprint arXiv:2406.07003},
  year={2024}
}

@inproceedings{wu2024repoformer,
  title={REPOFORMER: selective retrieval for repository-level code completion},
  author={Wu, Di and Ahmad, Wasi Uddin and Zhang, Dejiao and Ramanathan, Murali Krishna and Ma, Xiaofei},
  booktitle={Proceedings of the 41st International Conference on Machine Learning},
  pages={53270--53290},
  year={2024}
}

@inproceedings{cheng2024draco,
  title={Dataflow-Guided Retrieval Augmentation for Repository-Level Code Completion},
  author={Cheng, Wei and Wu, Yuhan and Hu, Wei},
  booktitle={Proceedings of the 62nd Annual Meeting of the Association for Computational Linguistics (Volume 1: Long Papers)},
  pages={7957--7977},
  year={2024}
}

@article{bui2024rambo,
  title={Rambo: Enhancing rag-based repository-level method body completion},
  author={Bui, Tuan-Dung and Luu-Van, Duc-Thieu and Nguyen, Thanh-Phat and Nguyen, Thu-Trang and Nguyen, Son and Vo, Hieu Dinh},
  journal={arXiv preprint arXiv:2409.15204},
  year={2024}
}

@article{husein2025completionreview,
  title={Large language models for code completion: A systematic literature review},
  author={Husein, Rasha Ahmad and Aburajouh, Hala and Catal, Cagatay},
  journal={Computer Standards \& Interfaces},
  volume={92},
  pages={103917},
  year={2025},
  publisher={Elsevier}
}

@article{jiang2024completionsurvey,
  title={A survey on large language models for code generation},
  author={Jiang, Juyong and Wang, Fan and Shen, Jiasi and Kim, Sungju and Kim, Sunghun},
  journal={arXiv preprint arXiv:2406.00515},
  year={2024}
}

@inproceedings{van2023completionmodel,
  title={Enriching source code with contextual data for code completion models: An empirical study},
  author={van Dam, Tim and Izadi, Maliheh and van Deursen, Arie},
  booktitle={2023 IEEE/ACM 20th International Conference on Mining Software Repositories (MSR)},
  pages={170--182},
  year={2023},
  organization={IEEE}
}

@article{wei2022cot,
  title={Chain-of-thought prompting elicits reasoning in large language models},
  author={Wei, Jason and Wang, Xuezhi and Schuurmans, Dale and Bosma, Maarten and Xia, Fei and Chi, Ed and Le, Quoc V and Zhou, Denny and others},
  journal={Advances in neural information processing systems},
  volume={35},
  pages={24824--24837},
  year={2022}
}

@inproceedings{diao2024cot2,
  title={Active Prompting with Chain-of-Thought for Large Language Models},
  author={Diao, Shizhe and Wang, Pengcheng and Lin, Yong and Pan, Rui and Liu, Xiang and Zhang, Tong},
  booktitle={62nd Annual Meeting of the Association for Computational Linguistics, ACL 2024},
  pages={1330--1350},
  year={2024},
  organization={Association for Computational Linguistics (ACL)}
}

@inproceedings{wermelinger2023copilot,
  title={Using github copilot to solve simple programming problems},
  author={Wermelinger, Michel},
  booktitle={Proceedings of the 54th ACM Technical Symposium on Computer Science Education V. 1},
  pages={172--178},
  year={2023}
}

@inproceedings{guo2025intention,
  title={Intention is All You Need: Refining Your Code from Your Intention},
  author={Guo, Qi and Xie, Xiaofei and Liu, Shangqing and Hu, Ming and Li, Xiaohong and Bu, Lei},
  booktitle={2025 IEEE/ACM 47th International Conference on Software Engineering (ICSE)},
  pages={728--728},
  year={2025},
  organization={IEEE Computer Society}
}

@inproceedings{ruan2025specrover,
  title={SpecRover: Code Intent Extraction via LLMs},
  author={Ruan, Haifeng and Zhang, Yuntong and Roychoudhury, Abhik},
  booktitle={2025 IEEE/ACM 47th International Conference on Software Engineering (ICSE)},
  pages={617--617},
  year={2025},
  organization={IEEE Computer Society}
}

@article{chen2021humaneval,
  title={Evaluating large language models trained on code},
  author={Chen, Mark and Tworek, Jerry and Jun, Heewoo and Yuan, Qiming and Pinto, Henrique Ponde De Oliveira and Kaplan, Jared and Edwards, Harri and Burda, Yuri and Joseph, Nicholas and Brockman, Greg and others},
  journal={arXiv preprint arXiv:2107.03374},
  year={2021}
}

@inproceedings{lu1codexglue,
  title={CodeXGLUE: A Machine Learning Benchmark Dataset for Code Understanding and Generation},
  author={Lu, Shuai and Guo, Daya and Ren, Shuo and Huang, Junjie and Svyatkovskiy, Alexey and Blanco, Ambrosio and Clement, Colin and Drain, Dawn and Jiang, Daxin and Tang, Duyu and others},
  booktitle={Thirty-fifth Conference on Neural Information Processing Systems Datasets and Benchmarks Track (Round 1)}
}

@inproceedings{jiang2023repair,
  title={Impact of code language models on automated program repair},
  author={Jiang, Nan and Liu, Kevin and Lutellier, Thibaud and Tan, Lin},
  booktitle={2023 IEEE/ACM 45th International Conference on Software Engineering (ICSE)},
  pages={1430--1442},
  year={2023},
  organization={IEEE}
}

@article{ren2020codebleu,
  title={Codebleu: a method for automatic evaluation of code synthesis},
  author={Ren, Shuo and Guo, Daya and Lu, Shuai and Zhou, Long and Liu, Shujie and Tang, Duyu and Sundaresan, Neel and Zhou, Ming and Blanco, Ambrosio and Ma, Shuai},
  journal={arXiv preprint arXiv:2009.10297},
  year={2020}
}

@article{hurst2024gpt,
  title={Gpt-4o system card},
  author={Hurst, Aaron and Lerer, Adam and Goucher, Adam P and Perelman, Adam and Ramesh, Aditya and Clark, Aidan and Ostrow, AJ and Welihinda, Akila and Hayes, Alan and Radford, Alec and others},
  journal={arXiv preprint arXiv:2410.21276},
  year={2024}
}

@article{zhu2024deepseekcoder,
  title={DeepSeek-Coder-V2: Breaking the Barrier of Closed-Source Models in Code Intelligence},
  author={Zhu, Qihao and Guo, Daya and Shao, Zhihong and Yang, Dejian and Wang, Peiyi and Xu, Runxin and Wu, Y and Li, Yukun and Gao, Huazuo and Ma, Shirong and others},
  journal={CoRR},
  year={2024}
}

@article{liu2024deepseekv3,
  title={Deepseek-v3 technical report},
  author={Liu, Aixin and Feng, Bei and Xue, Bing and Wang, Bingxuan and Wu, Bochao and Lu, Chengda and Zhao, Chenggang and Deng, Chengqi and Zhang, Chenyu and Ruan, Chong and others},
  journal={arXiv preprint arXiv:2412.19437},
  year={2024}
}

@article{roziere2023codellama,
  title={Code llama: Open foundation models for code},
  author={Roziere, Baptiste and Gehring, Jonas and Gloeckle, Fabian and Sootla, Sten and Gat, Itai and Tan, Xiaoqing Ellen and Adi, Yossi and Liu, Jingyu and Sauvestre, Romain and Remez, Tal and others},
  journal={arXiv preprint arXiv:2308.12950},
  year={2023}
}

@inproceedings{izadi2022codefill,
  title={Codefill: Multi-token code completion by jointly learning from structure and naming sequences},
  author={Izadi, Maliheh and Gismondi, Roberta and Gousios, Georgios},
  booktitle={Proceedings of the 44th international conference on software engineering},
  pages={401--412},
  year={2022}
}

@article{jiang2024aixcoder,
  title={aiXcoder-7B: A Lightweight and Effective Large Language Model for Code Processing},
  author={Jiang, Siyuan and Li, Jia and Zong, He and Liu, Huanyu and Zhu, Hao and Hu, Shukai and Li, Erlu and Ding, Jiazheng and Han, Yu and Ning, Wei and others},
  journal={arXiv preprint arXiv:2410.13187},
  year={2024}
}

@article{lewis2020rag,
  title={Retrieval-augmented generation for knowledge-intensive nlp tasks},
  author={Lewis, Patrick and Perez, Ethan and Piktus, Aleksandra and Petroni, Fabio and Karpukhin, Vladimir and Goyal, Naman and K{\"u}ttler, Heinrich and Lewis, Mike and Yih, Wen-tau and Rockt{\"a}schel, Tim and others},
  journal={Advances in neural information processing systems},
  volume={33},
  pages={9459--9474},
  year={2020}
}

@inproceedings{liurepobench,
  title={RepoBench: Benchmarking Repository-Level Code Auto-Completion Systems},
  author={Liu, Tianyang and Xu, Canwen and McAuley, Julian},
  booktitle={The Twelfth International Conference on Learning Representations}
}

@inproceedings{diao2024activecot,
  title={Active Prompting with Chain-of-Thought for Large Language Models},
  author={Diao, Shizhe and Wang, Pengcheng and Lin, Yong and Pan, Rui and Liu, Xiang and Zhang, Tong},
  booktitle={62nd Annual Meeting of the Association for Computational Linguistics, ACL 2024},
  pages={1330--1350},
  year={2024},
  organization={Association for Computational Linguistics (ACL)}
}

@inproceedings{zhao2024enhancing,
  title={Enhancing Zero-Shot Chain-of-Thought Reasoning in Large Language Models through Logic},
  author={Zhao, Xufeng and Li, Mengdi and Lu, Wenhao and Weber, Cornelius and Lee, Jae Hee and Chu, Kun and Wermter, Stefan},
  booktitle={Proceedings of the 2024 Joint International Conference on Computational Linguistics, Language Resources and Evaluation (LREC-COLING 2024)},
  pages={6144--6166},
  year={2024}
}

@article{li2025scodecot,
  title={Structured chain-of-thought prompting for code generation},
  author={Li, Jia and Li, Ge and Li, Yongmin and Jin, Zhi},
  journal={ACM Transactions on Software Engineering and Methodology},
  volume={34},
  number={2},
  pages={1--23},
  year={2025},
  publisher={ACM New York, NY}
}

@article{zhu2025uncertaintycot,
  title={Uncertainty-Guided Chain-of-Thought for Code Generation with LLMs},
  author={Zhu, Yuqi and Li, Ge and Jiang, Xue and Li, Jia and Mei, Hong and Jin, Zhi and Dong, Yihong},
  journal={CoRR},
  year={2025}
}

@inproceedings{li2024chainofcode,
  title={Chain of Code: Reasoning with a Language Model-Augmented Code Emulator},
  author={Li, Chengshu and Liang, Jacky and Zeng, Andy and Chen, Xinyun and Hausman, Karol and Sadigh, Dorsa and Levine, Sergey and Fei-Fei, Li and Xia, Fei and Ichter, Brian},
  booktitle={International Conference on Machine Learning},
  pages={28259--28277},
  year={2024},
  organization={PMLR}
}

@article{huang2023codecot,
  title={Codecot: Tackling code syntax errors in cot reasoning for code generation},
  author={Huang, Dong and Bu, Qingwen and Qing, Yuhao and Cui, Heming},
  journal={arXiv preprint arXiv:2308.08784},
  year={2023}
}

@article{yang2024codedebug,
  title={Enhancing the Code Debugging Ability of LLMs via Communicative Agent Based Data Refinement},
  author={Yang, Weiqing and Wang, Hanbin and Liu, Zhenghao and Li, Xinze and Yan, Yukun and Wang, Shuo and Gu, Yu and Yu, Minghe and Liu, Zhiyuan and Yu, Ge},
  journal={CoRR},
  year={2024}
}

@article{yang2024coast,
  title={Coast: Enhancing the code debugging ability of llms through communicative agent based data synthesis},
  author={Yang, Weiqing and Wang, Hanbin and Liu, Zhenghao and Li, Xinze and Yan, Yukun and Wang, Shuo and Gu, Yu and Yu, Minghe and Liu, Zhiyuan and Yu, Ge},
  journal={arXiv preprint arXiv:2408.05006},
  year={2024}
}

@inproceedings{nguyen2019function,
  title={Focus: A recommender system for mining api function calls and usage patterns},
  author={Nguyen, Phuong T and Di Rocco, Juri and Di Ruscio, Davide and Ochoa, Lina and Degueule, Thomas and Di Penta, Massimiliano},
  booktitle={2019 IEEE/ACM 41st International Conference on Software Engineering (ICSE)},
  pages={1050--1060},
  year={2019},
  organization={IEEE}
}

@article{hu2022lora,
  title={Lora: Low-rank adaptation of large language models.},
  author={Hu, Edward J and Shen, Yelong and Wallis, Phillip and Allen-Zhu, Zeyuan and Li, Yuanzhi and Wang, Shean and Wang, Lu and Chen, Weizhu and others},
  journal={ICLR},
  volume={1},
  number={2},
  pages={3},
  year={2022}
}

@article{kingma2014adam,
  title={Adam: A method for stochastic optimization},
  author={Kingma, Diederik P and Ba, Jimmy},
  journal={arXiv preprint arXiv:1412.6980},
  year={2014}
}

@incollection{singh1996stratified,
  title={Stratified sampling},
  author={Singh, Ravindra and Mangat, Naurang Singh},
  booktitle={Elements of survey sampling},
  pages={102--144},
  year={1996},
  publisher={Springer}
}

@inproceedings{wang2023self-instruct,
  title={Self-instruct: Aligning language models with self-generated instructions},
  author={Wang, Yizhong and Kordi, Yeganeh and Mishra, Swaroop and Liu, Alisa and Smith, Noah A and Khashabi, Daniel and Hajishirzi, Hannaneh},
  booktitle={Proceedings of the 61st annual meeting of the association for computational linguistics (volume 1: long papers)},
  pages={13484--13508},
  year={2023}
}

@article{luo2023wizardcoder,
  title={Wizardcoder: Empowering code large language models with evol-instruct},
  author={Luo, Ziyang and Xu, Can and Zhao, Pu and Sun, Qingfeng and Geng, Xiubo and Hu, Wenxiang and Tao, Chongyang and Ma, Jing and Lin, Qingwei and Jiang, Daxin},
  journal={arXiv preprint arXiv:2306.08568},
  year={2023}
}

@article{ahmad2025opencodeinstruct,
  title={OpenCodeInstruct: A Large-scale Instruction Tuning Dataset for Code LLMs},
  author={Ahmad, Wasi Uddin and Ficek, Aleksander and Samadi, Mehrzad and Huang, Jocelyn and Noroozi, Vahid and Majumdar, Somshubra and Ginsburg, Boris},
  journal={arXiv preprint arXiv:2504.04030},
  year={2025}
}

@article{chen2024reinstruct,
  title={Reinstruct: Building instruction data from unlabeled corpus},
  author={Chen, Shu and Guan, Xinyan and Lu, Yaojie and Lin, Hongyu and Han, Xianpei and Sun, Le},
  journal={arXiv preprint arXiv:2408.10663},
  year={2024}
}

\appendix

\end{document}